\documentclass[twocolumn, a4paper]{revtex4}

\usepackage[margin=0.8in]{geometry}
\usepackage{graphicx}
\usepackage{dcolumn}
\usepackage{bm}
\usepackage{setspace}
\usepackage{color}
\usepackage{sidecap}
\usepackage{amsmath}
\usepackage{titlesec}
\usepackage{multirow}

\titlespacing*{\section}{0pt}{1.1\baselineskip}{\baselineskip}

\begin{document}
\begin{singlespace}

\title{A model of cell-wall dynamics during sporulation in \textit{Bacillus subtilis}}

\author{Li-Wei Yap\textit{$^{1, \ddag}$} and Robert G. Endres\textit{$^{1, 2, *}$}}
\affiliation{
	{${}^1$Department of Life Sciences, Imperial College, London, United Kingdom.}\\
	{${}^2$Centre for Integrative Systems Biology and Bioinformatics, Imperial College, London, United Kingdom.}}

\begin{abstract}
	To survive starvation, \textit{Bacillus subtilis} forms durable spores. After asymmetric cell division, the septum grows around the forespore in a process called engulfment, but the mechanism of force generation is unknown. Here, we derived a novel biophysical model for the dynamics of cell-wall remodeling during engulfment based on a balancing of dissipative, active, and mechanical forces. By plotting phase diagrams, we predict that sporulation is promoted by a line tension from the attachment of the septum to the outer cell wall, as well as by an imbalance in turgor pressures in the mother-cell and forespore compartments. We also predict that significant mother-cell growth hinders engulfment. Hence, relatively simple physical principles may guide this complex biological process.
\end{abstract}

\maketitle

\section*{INTRODUCTION}
\textit{Bacillus subtilis} is a rod-shaped bacterium with a thick ($30$-$40$ nm) outer cell wall made of peptidoglycan (PG) polymers for withstanding high (${\sim}1.5$ MPa) turgor pressures \cite{banerjee}. To survive starvation, this bacterium forms robust and dormant endospores in several steps (Fig. 1): during DNA replication, septation is initiated asymmetrically by FtsZ (a), followed by pumping of one of the DNA molecules through the forespore's closing septum by ATP hydrolysis (b) \cite{errington, bisson, levine, narula}. Subsequently, the septum (made of PG) is remodeled and grows around the forespore (c), allowing the mother cell to engulf the forespore by its membrane for spore maturation (d) \cite{tocheva, meyer, doan}. This remodeling process is highly complicated, involving penicillin binding proteins (PBPs) to synthesize PG \cite{typas}, the PG-degradation enzymes SpoIID/M/P \cite{ojkic}, the SpoIIQ-SpoIIIAH backup mechanism \cite{ojkic}, and many other proteins \cite{meeske}. While MreB may help localise PBPs to the leading edge of engulfment, there are no known cytoskeletal force generators or motor proteins involved \cite{abanesdm}. What then drives PG remodeling and hence engulfment? We hypothesize that physical organizing principles may guide the engulfment process.

\begin{figure}[t]
	\centering
	\includegraphics[height=3.3cm]{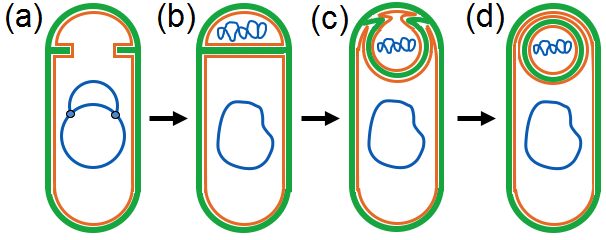}
	\caption{\label{fig:fig1} Schematic representation of morphological changes that occur during sporulation in \textit{Bacillus subtilis}. The cell wall is depicted in green, cell membranes in yellow, and DNA in blue. During DNA replication, septation is initiated near one of the poles (a), and the daughter DNA strand is pumped into the forespore (b). Then, the septum grows around the forespore (c), allowing the mother-cell membrane to engulf it (d).}
\end{figure}

Here, we aimed to derive a theoretical model of engulfment in the presence of cell-wall remodeling. To understand the various forces acting on the cell wall, we built on a theoretical framework recently introduced for studying the cell-wall dynamics at a cellular scale \cite{banerjee}. This framework is based on Rayleigh's principle of least-energy dissipation \cite{rayleigh} (equivalent to maximizing the rate of entropy production), with PG remodeling described as an active force that arises from changes in the mechanochemical energy associated with maintaining cell shape. We adapted it to mathematically derive a model for the dynamics of cell-wall remodeling during engulfment. By plotting comprehensive phase diagrams, we were able to determine the impact of various parameters on engulfment. Specifically, we predict that sporulation is driven by a positive line tension and an imbalance in turgor pressure between the mother cell and the forespore. We also investigated the theoretical relationship between engulfment and mother-cell growth, finding that significant growth hinders engulfment.

\section*{GENERAL EQUATION FOR CELL-WALL DYNAMICS}
Similar to the original framework \cite{banerjee}, the active and mechanical forces in our model are given by the derivatives of the total energy $E$ with respect to $N$ shape degrees of freedom. Similar to Ref. 1, we focused on the cell wall, but in the ESI\dag\, we also considered the role of the membrane \cite{pedrido} (Figs. S1a-b in ESI\dag). These shape degrees of freedom are specified by the generalized coordinates $q_i$ ($i = 1, ..., N$) and their respective velocities $\dot{q_i}$:
\setlength{\belowdisplayskip}{3pt} \setlength{\belowdisplayshortskip}{3pt}
\setlength{\abovedisplayskip}{3pt} \setlength{\abovedisplayshortskip}{3pt}
\begin{equation}
\eta_iV_i\frac{\dot{q_i}}{(q_i)^2} = -\frac{\partial{E}}{\partial{q_i}}, \qquad \forall{i} \label{eq:1}
\end{equation}
\noindent with viscosity constant $\eta_i$ and volume $V_i = h \cdot A_i$  over which dissipation of $q_i$ occurs. Here, $h$ is the thickness of the cell wall assumed to be constant and $A_i$ is the surface area of dissipation.

The left-hand side of Eq. \ref{eq:1} describes the dissipative force, whose corresponding energy represents the work done to the medium when the cell shape deforms at a rate $\dot{q_i}$, arising from the insertion of newly synthesized PG strands into the cell wall \cite{jiang10, jiang, jiang_rev}. The right-hand side describes the sum of active and mechanical forces, both of which, when integrated, represent the work done to the cell wall when the cell shape deforms by $q_i$. The active forces arise from distributed macromolecules that convert chemical energy into mechanical work; these forces include the chemical potential for PG synthesis, as well as the line tension caused by the active remodeling of the septum, which provides room for the mother-cell cytoplasm to entropically expand and hence engulf the forespore \cite{ojkic16}. The mechanical forces arise from the internal turgor pressure and the opposing surface tension of the elastic cell wall that act to increase and decrease cell volume, respectively. Further included in the mechanical force is the bending stiffness, reducing the degree of bending away from zero or any preferred curvature. We neglected the energy of interaction with cytoskeletal filaments due to lack of concrete experimental evidence. (The FtsZ-initiated septum is our initial condition in the models, and there are no established roles of FtsZ and MreB during engulfment.) Assuming no external forces or thermal noise, the dissipative energy is balanced by the active and mechanical energies \cite{rayleigh}. We implemented both a minimal model for engulfment and a more realistic model, which accounts for mother-cell and forespore growth.

\section*{MINIMAL MODEL OF ENGULFMENT}
\textit{B. subtilis} is modeled as a cylindrical cell with two hemispherical poles. Here, we use a minimal model for the purpose of gaining intuition. In this minimal model, we assumed for simplicity that the septum is already curved from the start (Fig. 2), although in reality the septum is initially flat \cite{tocheva}. Since the shape of the septum is fixed, the forespore is always spherical. Therefore, the radius $r$ and length $L$ of the central cylindrical region of the mother cell, as well as the different turgor pressures and volumes of the mother cell ($p_m$, and $V_m = \pi r^2L$) and forespore ($p_s$, and $V_s = 4\pi r^3/3$) are constant. The only shape degree of freedom is angle $\theta$ of engulfment (Fig. 2), and the surface area over which dissipation occurs during increase in $\theta$ is $A_\theta = 2\pi r^2 \sin{\theta}$. Surface area $A_s = 2\pi r^2 (1 + \sin{\theta})$ of the forespore cell wall and distance $r_s = r\cdot\cos{\theta}$ between the leading edge of the engulfing membrane and the longitudinal axis are both functions of $\theta$. The Helfrich bending energy of the septum $E^{bend}_s = k_s [2\pi r^2 (2/r)^2 + 2\pi r^2 \sin{\theta} (2/r - 2/R_0)^2]/2$ is also a function of $\theta$. Conversely, surface area $A_m = 4\pi r^2 + 2\pi rL$ of the mother-cell wall and bending energy of the mother-cell wall $E^{bend}_m = k_{m}[2\pi rL(1/r - 1/R_0)^2 + 4\pi r^2 (2/r - 2/R_0)^2]/2$ are not functions of $\theta$ and hence are constant. The bending energies are described in terms of circumferential bending rigidity ($k_m$ for mother cell, $k_s$ for forespore) and preferred radius $R_0$ of the cell-wall cross-section.

\begin{figure}[t]
	\centering
	\includegraphics[height=6.2cm]{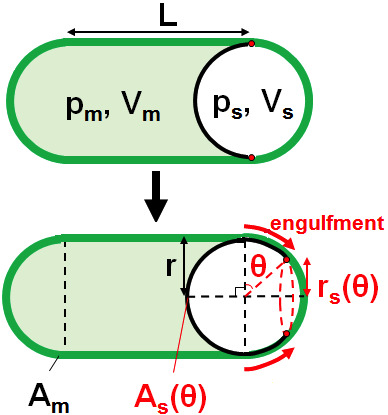}
	\caption{\label{fig:fig2} Minimal model of engulfment. The only shape degree of freedom is angle $\theta$ of engulfment. The red dots represent the leading edge of the engulfing membrane, and $\theta$ increases over time to maximally $\pi/2$. Surface area of the forespore cell wall, $A_s$, as well as distance between the leading edge and the longitudinal axis, $r_s$, change in response to $\theta$. All other parameters including cell radius $r$, cell length $L$, pressure $p$, volume $V$, and the surface area $A_m$ of the mother-cell wall are constant.}
\end{figure}

The sum of active and mechanical energies is given by:
\setlength{\belowdisplayskip}{3pt} \setlength{\belowdisplayshortskip}{3pt}
\setlength{\abovedisplayskip}{3pt} \setlength{\abovedisplayshortskip}{3pt}
\begin{multline}
E = - p_mV_m - p_sV_s + (\gamma - \varepsilon)(A_m + A_s) + 2\pi r_sf \\ + E^{bend}_m + E^{bend}_s, \label{eq:2}
\end{multline}
with surface tension $\gamma$, chemical potential $\varepsilon$ for PG remodeling, line tension $f$, mother-cell and forespore turgor pressures $p_m$ and $p_s$, and cell-wall and septum bending energies $E^{bend}_m$ and $E^{bend}_s$. Similar to Ref. 1, we used a constant surface tension as $r$ is either fixed or strongly constrained by MreB so that different functional forms of $\gamma$ would not have much effect. The line tension may represent the energy cost for remodeling the attachment of the septum to the outer cell wall by the SpoIID/M/P complex \cite{ojkic, ojkic16}, or originate from the membrane, which has to bend backwards onto itself. This expression for $E$ was substituted into Eq. \ref{eq:1} for cell-wall dynamics with $q_1 = \theta$. Since we are interested in the partial derivative of $E$ with respect to $\theta$, the terms of Eq. (2) that are either constant or not functions of $\theta$ can be ignored:
\begin{equation}
E \approx \delta \cdot A_s + 2\pi f \cdot r_s \,+\,E^{bend}_s, \label{eq:3}
\end{equation}
with $\delta = \gamma - \varepsilon$. Using Eq. (3) in Eq. (1), we obtained:
\begin{multline}
\frac{d\theta}{dt} =  \frac{\mu_\theta \cdot \theta^2}{r^2 \cdot \sin{\theta}} \cdot \Big[2\pi rf \cdot\sin{\theta} - \delta \cdot 2\pi r^2\cdot \cos{\theta} \\ - 4k_s\pi r^2 \Big(\frac{1}{r} - \frac{1}{R_0}\Big)^2 \cdot\cos{\theta}\Big], \label{eq:4}
\end{multline}
where $\mu_\theta = 1/(2\pi h\eta_\theta)$ is the mobility coefficient of engulfment. To make the various parameters dimensionless, surface tension was rescaled as $\tilde{\gamma} = \gamma/(pR_0)$, chemical potential as $\tilde{\varepsilon} = \varepsilon/(pR_0)$, line tension as $\tilde{f} = f/(pR_0^2)$, and circumferential bending rigidity of the mother-cell and forespore as $\tilde{k}_m = k_m/(pR_0^3) = 3.6$ \cite{banerjee} and $\tilde{k}_s = k_s/(pR_0^3) = 0.18$ (see Table S1 in ESI\dag) respectively, 
with $p = 1.5$ MPa \cite{banerjee, lan}, $R_0 = 0.43$ $\mu$m \cite{banerjee} and $\tilde{\delta} = \tilde{\gamma} - \tilde{\varepsilon}$. Since $r$ is fixed at the same value as $R_0$ (see Table S1 in ESI\dag), the last term $-4k_s\pi r^2 \Big(\frac{1}{r} - \frac{1}{R_0}\Big)^2 \cdot\cos{\theta}$ in Eq. (4) effectively cancels out. Moreover, as the septum and forespore cell wall are initially assumed to be a single PG layer \cite{ojkic16}, $k_s \ll k_m$, so $E^{bend}_s$ has a minor contribution to the engulfment dynamics (Fig. S1f in ESI\dag). As $\tilde{f}$ and $\tilde{\delta}$ are not well-constrained by experiments, we scanned through these parameters.

We first investigated the conditions of $\tilde{\delta}$ and $\tilde{f}$ that favor engulfment. For engulfment to occur, energy must be released into the environment, i.e. $\partial{E}/\partial{\theta} < 0$, allowing $\theta$ to increase from $0$ to the maximum $\pi/2$. As evident from Eq. \ref{eq:3}, engulfment is favored when $\tilde{\delta} < 0$, i.e. when the chemical potential for PG remodeling is greater than the surface tension (Fig. 3a). Thus, there is competition between $\tilde{\gamma}$ and $\tilde{\varepsilon}$, where $\tilde{\varepsilon}$ favors engulfment, whereas $\tilde{\gamma}$ represents an energy penalty for engulfment (both parameters are assumed positive). The more negative $\tilde{\delta}$ is, the easier it is to overcome the energy barrier for engulfment. Engulfment is favored when $\tilde{f} > 0$ (Fig. 3b). This is because $\tilde{f}$ is multiplied by $r_s = r\cos{\theta}$, and $r_s$ decreases as $\theta$ increases and engulfment proceeds. In fact, if $\tilde{f}$ is sufficiently large, engulfment may occur even for positive $\tilde{\delta}$ (Fig. 3c). It may seem strange that engulfment is favored by a positive line tension $\tilde{f}$, because a tension represents an energy penalty. However, engulfment forces are about changes in energy, and engulfment reduces the penalty from the line tension due to decreasing radius $r_s$. Hence, Figs. 3a-b show that engulfment is driven by both growth ($\tilde{\delta} < 0$) and line tension ($\tilde{f} > 0$). In the ESI\dag, we also varied $r$, and found that deviation from the preferred radius $R_0$ has limited effect on the plot for $\theta(t)$ (Fig. S1c in ESI\dag).

\begin{figure}[t]
	\centering
	\includegraphics[height=13cm]{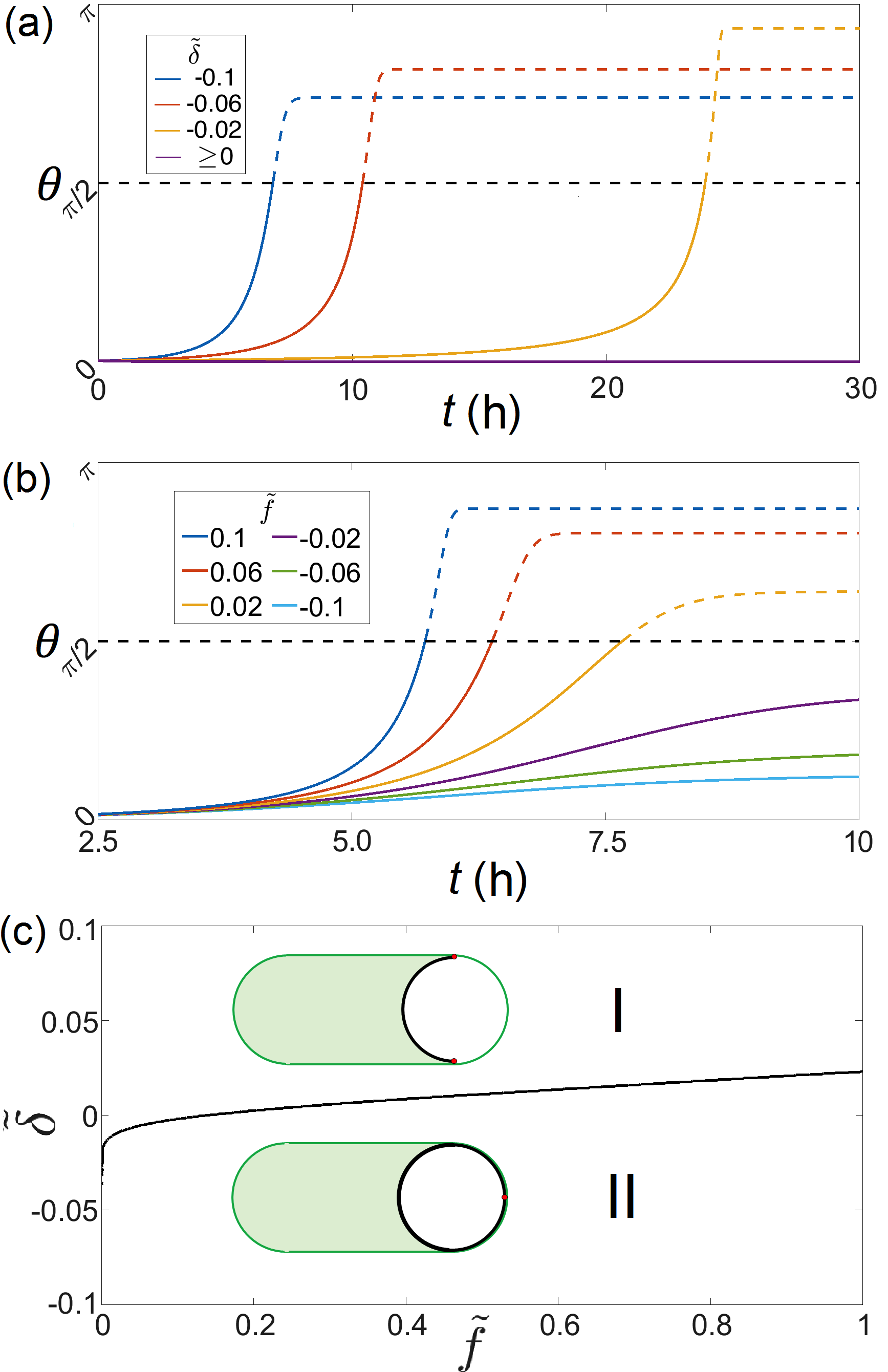}
	\caption{\label{fig:fig3} Ensemble plots and phase diagram for minimal model. (a) Plot of $\theta(t)$ when difference $\tilde{\delta} = \tilde{\gamma} - \tilde{\varepsilon}$ between surface tension and chemical potential is varied, whilst line tension $\tilde{f} = 0.04$ is kept constant. The solid lines represent engulfment up to $\pi/2$ (engulfment completed), whereas the dashed lines represent dynamics beyond $\pi/2$. Engulfment is favored for $\tilde{\delta} < 0$. (b) Plot of $\theta(t)$ when $\tilde{f}$ is varied, whilst $\tilde{\delta} = -0.1$ is kept constant. Engulfment is favored for $\tilde{f} > 0$. (c) Phase diagram in the ($\tilde{\delta}$, $\tilde{f}$) plane. For all other parameters, see Table S1 in ESI\dag.}
\end{figure}

We analytically verified the steady-state angles of engulfment $\theta^*$ (Figs. 3a-b) using Eq. \ref{eq:4}. The lower steady state ($\theta^*_1 = 0$) is in fact the initial $\theta(t=0)$, which increases over time towards the upper steady state ($\theta^*_2 = \tan^{-1}{[(r\tilde{\delta} + 2r\tilde{k}_{s}(1/r - 1/R_0)^2)/\tilde{f}]}$). Although engulfment is completed when $\theta = \pi/2$, the upper steady state might be greater than $\pi/2$. In fact, $\tan{(\pi/2)}$ is undefined, so only in the absence of line tension ($\tilde{f} = 0$) will the steady state be exactly $\pi/2$. This implies excess energy in the cell wall when engulfment is complete. The larger the difference between $\pi/2$ and the upper steady state, the higher the engulfment rate at $\pi/2$ (Figs. 3a-b). This excess energy might be used to promote membrane fission \cite{doan}.

Linear stability analysis was conducted for all values of $\tilde{\delta}$ and $\tilde{f}$ in the ensemble plots (Figs. 3a-b) to determine stability at both steady states $\theta^*_{1,2}$ using $\dot{\delta\theta}\simeq g'(\theta^*)\delta\theta$. If $g'(\theta^*) > 0$, the perturbation ($\delta\theta$) grows exponentially, indicating unstable equilibrium, whereas if $g'(\theta^*) < 0$, $\delta\theta$ dampens out, indicating stable equilibrium \cite{strogatz}. Note that $g'(\theta^* = 0) = 0$, which indicates a need for energy consumption at $t = 0$. We found that there is instability at small $\theta$ (near $\theta_1^*$) and stability at $\theta_2^*$, which explains the increase in $\theta$ over time. To analyze the stability of the lower steady state $\theta_1^* = 0$ further, we take the limit of $\theta\searrow0$ in Eq. \ref{eq:4}, leading to $d\theta/dt = -2\pi\mu_\theta\tilde{\delta}\cdot\theta$, where engulfment proceeds for $\tilde{\delta} < 0$, i.e. $\tilde{\varepsilon} > \tilde{\gamma}$. Hence, synthesis is required to get engulfment started.

The mother cell synthesizes considerable amounts of membrane, also required for compartment-specific expression of transcription factors \cite{pedrido, tan}. To study the effect of membrane synthesis on the cell wall, we extended the minimal model in the ESI\dag\, to include the membrane surface areas (Fig. S1b in ESI\dag). Assuming that the chemical potentials for synthesizing the cell wall and membrane are the same, the value of $\tilde{\delta}$ in the ordinate axis of the phase diagram (Fig. 3c) is effectively reduced, so that it is easier for cells to be in Region II in which engulfment occurs. Whilst this seems counter-intuitive as additional membrane synthesis is required (a cost), we assumed that there is sufficient energy available to drive membrane synthesis. This implies that engulfment actually relieves this drive or \textquoteleft pressure'.

\section*{REALISTIC MODEL OF ENGULFMENT}
In reality the septum is initially flat, so that the forespore is hemispherical prior to engulfment \cite{tocheva, meyer}. Over time, the septum becomes increasingly curved as the forespore expands into the mother cell to form a hemispheroid joined to the initial hemisphere. Hence, there might be competition between engulfment and forespore expansion for limited resources during starvation.

\begin{figure}[t]
	\centering
	\includegraphics[height=7cm]{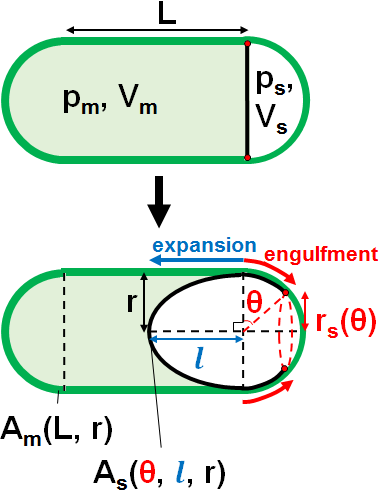}
	\caption{Realistic model of engulfment. There are two main shape degrees of freedom; angle $\theta$ of engulfment and expansion $l$ of the forespore into the mother cell. With mother cell growth during engulfment, there are two more shape degrees of freedom; cell radius $r$ and cell length $L$. The red dots represent the leading edge of the engulfing membrane, and $\theta$ increases over time to maximally $\pi/2$. Mother-cell and forespore volumes ($V_m$ and $V_s$) and cell-wall surface areas ($A_m$ and $A_s$), as well as distance between the leading edge and the longitudinal axis, $r_s$, are functions of one or more shape degrees of freedom.}
\end{figure}

With mother-cell growth, there are four shape degrees of freedom: angle $\theta$ of engulfment, expansion $l$ of the forespore into the mother cell along its longitudinal axis, as well as the mother-cell radius $r$ and length $L$ (Fig. 4). The mother cell and forespore volumes are $V_m = \frac{2}{3} \pi r^3 + \pi r^2 L - \frac{2}{3} \pi r^2 l$ and $V_s = \frac{2}{3} \pi r^2 (l + r)$, respectively. The forespore cell wall surface area is $A_s = \pi r^2 [1 + l\cdot \sin^{-1}(\sqrt{1 - r^2 / l^2}) / (r\sqrt{1 - r^2 / l^2}) + 2\sin{\theta}]$. The mother cell wall surface area $A_m$, distance $r_s$ between leading edge of engulfing membrane and longitudinal axis, the surface area $A_\theta$ over which dissipation occurs during increase in $\theta$, as well as the cell-wall bending energy $E^{bend}_m$, are the same as in the minimal model. Since $k_s \ll k_m$, $E^{bend}_s$ has a minor contribution to the engulfment dynamics, we neglect $E^{bend}_s$ in order to remain having analytical expressions for the energies (further explained in Discussion and Conclusions). The surface area over which dissipation occurs during increase in $l$ is $A_l = \pi r^2 [1 + l\cdot \sin^{-1}(\sqrt{1 - r^2 / l^2}) / (r\sqrt{1 - r^2 / l^2})]$. The surface area over which dissipation occurs during increase in $r$ is $A_r = A_m + A_s$. The surface area over which dissipation occurs during increase in $L$ is $A_L = 2\pi rL$. With Eq. (2) for the sum of active and mechanical energies, we derived $dl/dt$, $dr/dt$, and $dL/dt$ (see ESI\dag\, for the complete formulae).

We initially assumed for simplicity that no resources are diverted to mother-cell growth due to starvation, so $r$ and $L$ are constant. This is consistent with previously published time-lapse microscopy data \cite{meyer, ojkic16}, showing that the cell volume remains constant throughout engulfment. (Later, the mother cell is allowed to grow during sporulation, so that $r$ and $L$ increase simultaneously with $\theta$ and $l$.) To make the various parameters dimensionless, surface tension and circumferential bending rigidity were rescaled as in the minimal model. The initial radius $r_0$ is set to $R_0$ \cite{banerjee}, whilst the initial length $L_0$ was set to $3.4\,\mu$m, which is the average experimentally measured value \cite{sharpe}. 

\begin{figure}[t]
	\centering
	\includegraphics[height=8.75cm]{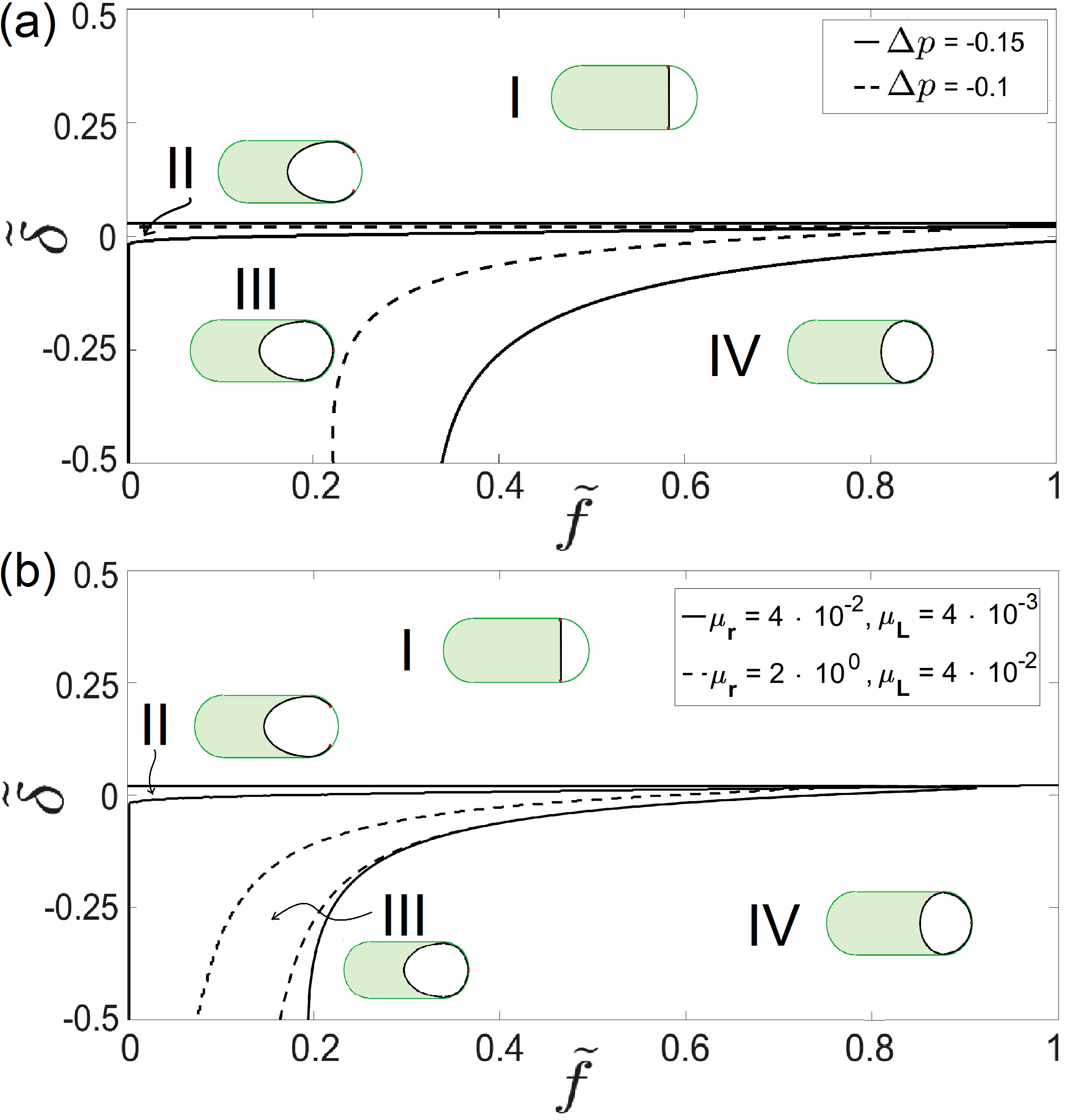}
	\caption{Phase diagrams in ($\tilde{\delta},\tilde{f}$) plane for realistic model. (a) Effect of $\Delta p$ in units of MPa for $\mu_r = \mu_L = 0$ (no growth of mother cell). As the mother-cell pressure increases relative to the forespore pressure, $\Delta p$ and Region IV increase, whilst Region III decreases. The lines in the phase diagram were determined by contour plots, which showed whether engulfment or expansion are complete for hundreds of thousands of combinations of parameters, including $\tilde{\delta}$, $\tilde{f}$, and $\Delta p$. The contour plots were then superimposed to obtain regions that show whether engulfment is completed before expansion. (b) Effect of $\mu_r$ and $\mu_L$ at fixed $\Delta p = -0.1$ MPa. In contrast to (a), limited growth leads to small increase in Region IV but significant growth leads to drastic increase in Region II and in turn drastic decrease in Region III. For all other parameters, see Table S1 in ESI\dag.}
\end{figure}

We investigated the conditions of turgor pressure that favor an increase in $\theta(t)$ and $l(t)$. Phase diagrams were plotted for varying values of the pressure difference between the mother cell and forespore $\Delta p = p_m - p_s$ (Fig. 5a). The value of $\tilde{f}$ was set to $0.2$, as in the original framework \cite{banerjee}, and we chose $\tilde{\delta} = -0.5$ \cite{banerjee}. In the phase diagrams, there are four different regions where engulfment and/or forespore expansion are favored, but these regions have different sizes depending on $\Delta p$. Region I represents the absence of both engulfment and forespore expansion. Region II represents the situation of incomplete engulfment (terminated at $L/2$) and excessive forespore expansion. Region III represents the situation where forespore expansion is completed before engulfment. Finally, Region IV represents the situation where forespore engulfment is completed without major forespore expansion.

\begin{figure}[t]
	\centering
	\includegraphics[height=9.1cm]{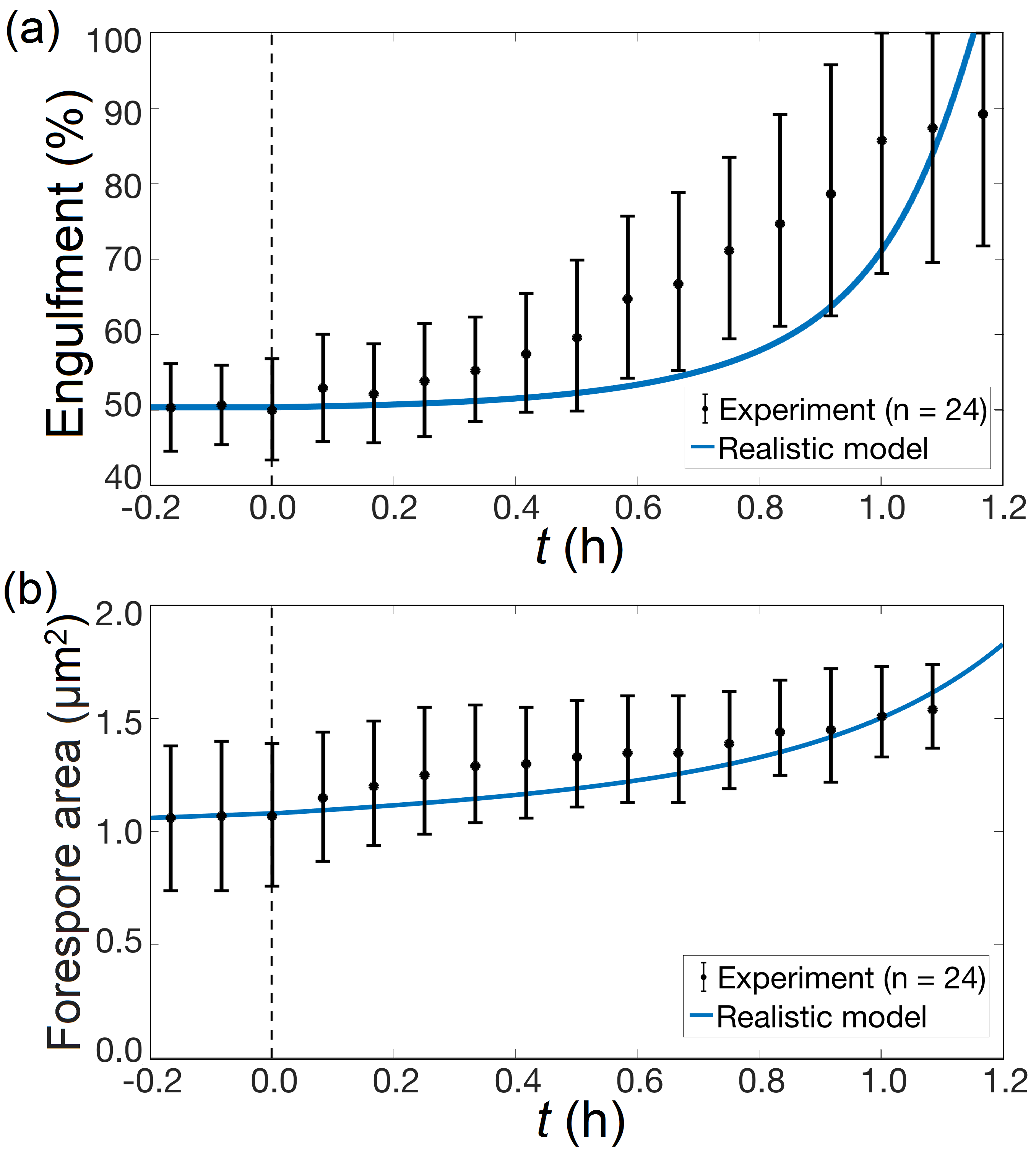}
	\caption{Comparison of model predictions with time-lapse microscopy data ($n$ is number of individual cells observed). (a) Time trace of experimentally measured engulfment \cite{ojkic16} (black symbols) and results from numerical calculation using realistic model (blue solid line). 50\% engulfment is assumed to occur as soon as septum formation is complete and prior to septum remodeling. The vertical dashed line represents the start of engulfment at $0.0\;h$. Parameters: $\mu_\theta = 1.2 \; m^2 J^{-1} h^{-1}$, $\tilde{\delta} = -0.5$, $\tilde{f} = 0.2$, and $\Delta p = -0.1$ MPa; for all other parameters, see Table S1 in ESI\dag. (b) Time trace of experimentally measured forespore surface area $A_s$ \cite{ojkic16} (black symbols) and results from numerical calculation using realistic model (blue solid line). Parameters: $\mu_\theta = 1.2 \; m^2 J^{-1} h^{-1}$, $r = 0.58 \; \mu m$, $\tilde{\delta} = -0.35$, $\tilde{f} = 0.2$, and $\Delta p = -0.1$ MPa; for all other parameters, see Table S1 in ESI\dag.}
\end{figure}

To understand which region in Fig. 5a might be physiologically relevant, we compared the model with time-lapse microscopy data \cite{ojkic16}. For $\tilde{\delta} < 0$, $\tilde{f} > 0$, and $\Delta p < 0$ MPa, our realistic model matched experimental measurements of time-dependent engulfment and forespore surface area $A_s$ (Figs. 6a-b, see caption as well as Table S1 in ESI\dag\;for extracted parameter values). The experimental data \cite{ojkic16} only shows little increase in $l$ once engulfment is complete, so Region III represents the ideal set of parameter values for engulfment and forespore expansion. As $\Delta p$ decreases, Region III increases, whilst Region IV decreases (transition from dashed to solid lines). This confirms a previous hypothesis that the forespore expands because of higher osmolarity and in turn higher turgor pressure than in the mother cell \cite{stragier}. Further support comes from the ensemble plot of $l(t)$ for $-0.3 \leq \Delta p \leq 0.3$ MPa (Fig. S2b in ESI\dag). The dashed lines in the ensemble plot show forespore expansion if allowed to continue after engulfment is complete, and for $\Delta p \geq 0$ MPa, $l(t)$ is predicted to increase sharply to the point of complete forespore expansion only after engulfment is complete. This, taken together with the experimental data \cite{ojkic16}, suggests that $\Delta p < 0$ MPa.

Linear stability analysis of $\theta$ again shows instability at small $\theta$ (near $\theta_1^*$) and stability at $\theta_2^* = \tan^{-1}{[(r\tilde{\delta} + 2r\tilde{k}_{s}(1/r - 1/R_0)^2)/\tilde{f}]}$, which explains the increase in $\theta$ over time. Linear stability analysis of $l$ shows that small $l$ (near $l^* = 0$) is unstable for $\Delta p < 0$ but stable for $\Delta p > 0$ MPa, so the sharp increase in $l(t)$ occurs earlier when $\Delta p < 0$ MPa (Fig. 3b). Interestingly, this sharp increase occurs even though the equilibrium is much more unstable at $l = L/2$ than at small $l$, suggesting a need for energy for forespore expansion. Perhaps forespore expansion is driven by the utilization of limited energy resources in the cytoplasm \cite{freese}. Furthermore, the instability at $l = L/2$ indicates excess energy in the cell wall, which might be used to promote forespore maturation \cite{eijlander}.

For fixed $r$ and $L$, we wondered how forespore shape influences engulfment. For that purpose, we compared the engulfment dynamics of the realistic (with fixed $r$ and $L$) and the minimal model (see ESI\dag). Indeed, as differing only by forespore shape, we found that both models yield nearly the same engulfment dynamics for the same set of parameter values, e.g. the same plot for $\theta(t)$ is produced by the minimal and the realistic model for $\tilde{\delta} = -0.5$ and $\tilde{f} = 0.2$ (Figs. S2a and S2c in ESI\dag).

Finally, we investigated whether engulfment and/or forespore expansion would still be favored if $r(t)$ and $L(t)$ are allowed to increase simultaneously with $\theta(t)$ and $l(t)$. Phase diagrams for different $\mu_r$ and $\mu_L$ were plotted for a constant $\Delta p$ in the ($\tilde{\delta},\tilde{f}$) plane, using the same values of $\mu_\theta$ and $\mu_l$ as before (Fig. 5b). Here, $\mu_r = 1/(\pi h\eta_r)$ and $\mu_L = 1/(2\pi h\eta_L)$ are the mobility coefficients of radial and longitudinal growth, respectively. Notably, if $\mu_r = \mu_L = 0$, then the phase diagram is the same as in Fig. 5a, since $dr/dt = dL/dt = 0$.

For large values of $\mu_r = 2 \; m^2 J^{-1} h^{-1}$ and $\mu_L = 4\cdot10^{-2} \; m^2 J^{-1} h^{-1}$, mother-cell growth is significant, leading to a significant increase in Region II and a significant decrease in Region III (Fig. 5b, dashed lines). Thus, there is a higher chance that engulfment is not completed. Hence, significant growth is detrimental to sporulation, potentially because a large amount of chemical potential is required for PG remodeling. If this chemical potential is diverted to mother-cell growth, then there would not be enough resources for engulfment or forespore expansion, in line with sporulation being a starvation response. However, for small values of $\mu_r = 4\cdot10^{-2} \; m^2 J^{-1} h^{-1}$ and $\mu_L = 4\cdot10^{-3} \; m^2 J^{-1} h^{-1}$, mother-cell growth is limited and leads to only a slight decrease in Region III and no increase in Region II (solid lines). Thus, the inhibition of engulfment by growth is limited.  Although rod-shaped bacteria like \textit{B. subtilis} usually grow longitudinally rather than radially \cite{angert}, our model predicts similar behavior of $r(\theta)$ and $L(\theta)$ as compared to the original framework \cite{banerjee}, especially for $\Delta p = -0.1$ MPa. This can be seen from the plot for $r(\theta)$, which increases slightly and reaches a plateau (Figs. S3c and S4c in ESI\dag), as well as from the plot for $L(\theta)$, which increases linearly (Figs. S3d and S4d in ESI\dag), whilst engulfment is not yet complete. That we re-obtain similar behavior of $r(t)$ and $L(t)$ is remarkable given that the original framework does not account for sporulation.

\section*{DISCUSSION AND CONCLUSIONS}
Elucidating the dynamics of cell-wall remodeling is key to understanding the conditions that favor engulfment during sporulation in \textit{B. subtilis}. With our model, we showed that it is energetically possible for PG remodeling to drive membrane migration and forespore expansion. Thus, PG likely arose before the evolution of sporulation as the latter is a consequence of PG biochemistry \cite{tocheva16}. To promote sporulation, PG remodeling is surprisingly aided by a line tension and a turgor-pressure imbalance in the mother cell and forespore compartments. Furthermore, significant mother-cell growth is detrimental to sporulation, as it diverts excessive energy resources from engulfment. How could these predictions be tested? Pressure could be measured in the mother cell and forespore by AFM indentation and fluorescence microscopy \cite{arnoldi, deng}. Indeed, packing of charged DNA in the small forespore may induce osmotic swelling in line with Region III \cite{garrido}. High-throughput imaging could be used to test if some cells grow despite sporulation.

While predictive, our modeling is minimalistic and a number of simplifications were necessarily made. First, similar to earlier models \cite{banerjee, jiang10, jiang, jiang_rev}, membrane energetics and dynamics were neglected. Indeed, as shown in Figs. S1a-b with the minimal model, membrane contributions are negligible (as long as there is sufficient energy available to drive membrane synthesis). Second, the detailed shape of the forespore cell wall is unknown. However, by comparing the spherical and spheroidal forespore shapes without mother-cell growth, the results from the ensemble plots for $\theta(t)$ are very similar without qualitative differences (Figs. S2a and S2c in ESI\dag). Third, the cell-wall bending energy of the forespore $E^{bend}_s$ was neglected in Section 3. As the septum and forespore are initially assumed to be a single PG layer \cite{ojkic16}, using a significantly smaller bending stiffness for the forespore in the minimal model shows that its contribution to engulfment is indeed minor, especially when $r(t)$ does not grow by more than 5\% (Fig. S1f in ESI\dag). Indeed, $r(t)$ increases by 5\% in the case of significant growth (Fig. S3c in ESI\dag) and by 3\% in the case of limited growth (Fig. S4c in ESI\dag), justifying this model simplification.

Another model simplification concerns how the cell-wall surface energy is described. In Eq. (2) we assumed a constant surface tension $\gamma$, with the area energy given by $\gamma \cdot A$, where $A$ is the total cell-wall area of mother cell and forespore. This models a plastic cell wall, determined by growth \cite{amir}. In contrast, elastic cell-wall deformations could be considered around the minimum-free energy surface area determined by Laplace law, where pressure, surface tension, and geometry are all connected. However, Laplace law does not lead to stable rod shapes \cite{koch}, and in sporulation, the pressures in mother cell and forespore are likely different, which would lead to different radii in mother cell and forespore compartments. However, this has not been observed \cite{ojkic16}. Furthermore, rod-shaped bacteria in the original framework \cite{banerjee} are modeled with a constant surface tension as well, leading to the same growth law as more detailed models with elastic strain \cite{jiang, jiang_rev}. Lastly, including a small non-linear elastic correction \cite{herant} $\gamma_0\cdot A^2$ to our minimal model ($\gamma_0/\gamma \leq 0.1$) has only minor effects on engulfment (Figs. S1d-e in ESI\dag). This is in line with previous studies involving microfluidics \cite{amir}, which showed that \textit{B. subtilis} cell-wall deformations during PG synthesis are usually plastic, whereas any elastic deformations tend to be transient and hence have little relation to growth. Taken together, our presented model provides insights while being minimal and conceptually straightforward to interpret.

We believe that our findings could aid the design of whole-cell radiation-biosensors, insecticides, probiotics, vectors for the delivery of drugs, vaccine antigens, or immunomodulators \cite{barak}. Furthermore, some spore-forming bacteria constitute major health threats, such as \textit{Clostridium difficile} and \textit{Bacillus anthracis} \cite{ojkic16}, and our model may provide insights into preventing spore formation e.g. by influencing mother-cell growth, turgor pressure, or line tension. Indeed, our findings are of great interest to soft matter, because bacterial cell walls, like other biological systems, exhibit properties rarely found in condensed matter physics which are often caused by growth \cite{amir}. Future work may incorporate the effects of stochasticity and molecular-scale defects for closer connection with molecular biology experiments \cite{banerjee, ojkic16}.

\section*{ACKNOWLEDGEMENTS}
We thank N. Ojkic for help with the modeling as well as J. L\'{o}pez-Garrido, G. Salbreux, and S. Banerjee for helpful discussions. R.G.E. acknowledges financial support from the European Research Council Starting-Grant No. 280492-PPHPI and the Biotechnology and Biological Sciences Research Council Grant No. BB/I019987/1.

\vspace{5mm}
\noindent $\ddag\ $Present address: Department of Biosystems Science and Engineering (D-BSSE), ETH Z{\"u}rich, Switzerland.  Email: liwei.yap13@alumni.imperial.ac.uk\\
\noindent $\ast\ $Email: r.endres@imperial.ac.uk\\
\noindent $\dagger\ $Electronic Supplementary Information (ESI) available. See DOI: 10.1039/b000000x/

\end{singlespace}
\end{document}